\documentclass[11pt]{article}

\usepackage{fullpage,latexsym,amsmath,amsfonts}

\usepackage{xcolor} 
\usepackage{graphicx}
\usepackage{authblk}

\def\01{\{0,1\}}

\newcommand{\ket}[1]{\left|#1\right\rangle}
\newcommand{\bra}[1]{\left\langle#1\right|}
\newcommand{\ketbra}[2]
{|#1\rangle\langle#2|}
\newcommand{\proj}[1]{\ketbra{#1}{#1}}
\newcommand{\braket}[2]{\langle#1|#2\rangle} 

\newtheorem{result}{Result}

\newcommand{\vs}{\underline{S}}

\newcommand{\hatZ}{\overset{\sqcap}{Z} }
\newcommand{\be}{\begin{equation}}
\newcommand{\ee}{\end{equation}}
\newcommand{\eff}{\mathrm{eff}}
\newcommand{\jmax}{s}
\newcommand{\jmin}{s-1}
\newcommand{\mj}{m}
\newcommand{\Hint}{V_{\textrm{SG}}}
\newcommand{\DeltaV}{\mathcal{D}(V)}

\newenvironment{proof}
{\noindent {\bf Proof. }}
{{\hfill $\Box$}\\
 \smallskip}

\begin{document}

\title{Can a spin-half particle ever give more than two spots in a Stern-Gerlach experiment? - the subtle physics of effective Hamiltonians.}

\author[1]{Noah Linden\thanks{n.linden@bristol.ac.uk}}
\author[2]{Sandu Popescu\thanks{s.popescu@bristol.ac.uk}}
\author[2]{Anthony J. Short\thanks{tony.short@bristol.ac.uk}}

\affil[1]{\small \emph{School of Mathematics, University of Bristol, Woodland Road, Bristol BS8 1UG, United Kingdom}} 
\affil[2]{\small \emph{H. H. Wills Physics Laboratory, University of Bristol, Tyndall Avenue, Bristol BS8 1TL, United Kingdom}}

\date{}
\maketitle

\begin{abstract}
We show that a spin-1/2 particle can behave as if it were spin-$s$, and generate 2s+1 spots in a Stern Gerlach measurement (albeit with a smaller gyromagnetic ratio).  This arises from some subtle properties of effective Hamiltonians and Hamiltonians with constraints.  Examples of implications of the effect in condensed matter are discussed. We also give some simple non-perturbative bounds for a system subjected to strong constraints.

\end{abstract}

\section{Introduction}

One of the most fundamental experimental results in the early development of quantum theory was the Stern-Gerlach experiment \cite{Gerlach1922}, in which an inhomogeneous magnetic field was used to measure the z-component of spin of a beam of spin-half particles, and two distinct spots were observed on the screen, corresponding to the two possible eigenvalues of the spin ($s_z = \pm \frac{\hbar}{2}$). But can spin-half particles being measured using a Stern-Gerlach experiment ever give more than two spots? In this paper, we will show the surprising result that they can. 

As we will show, this is a consequence of the subtle properties of effective Hamiltonians, and the ways in which they can affect the fundamental properties of a system, including their effective dimension. 

\begin{figure}
    \centering
    \includegraphics[width=15cm]{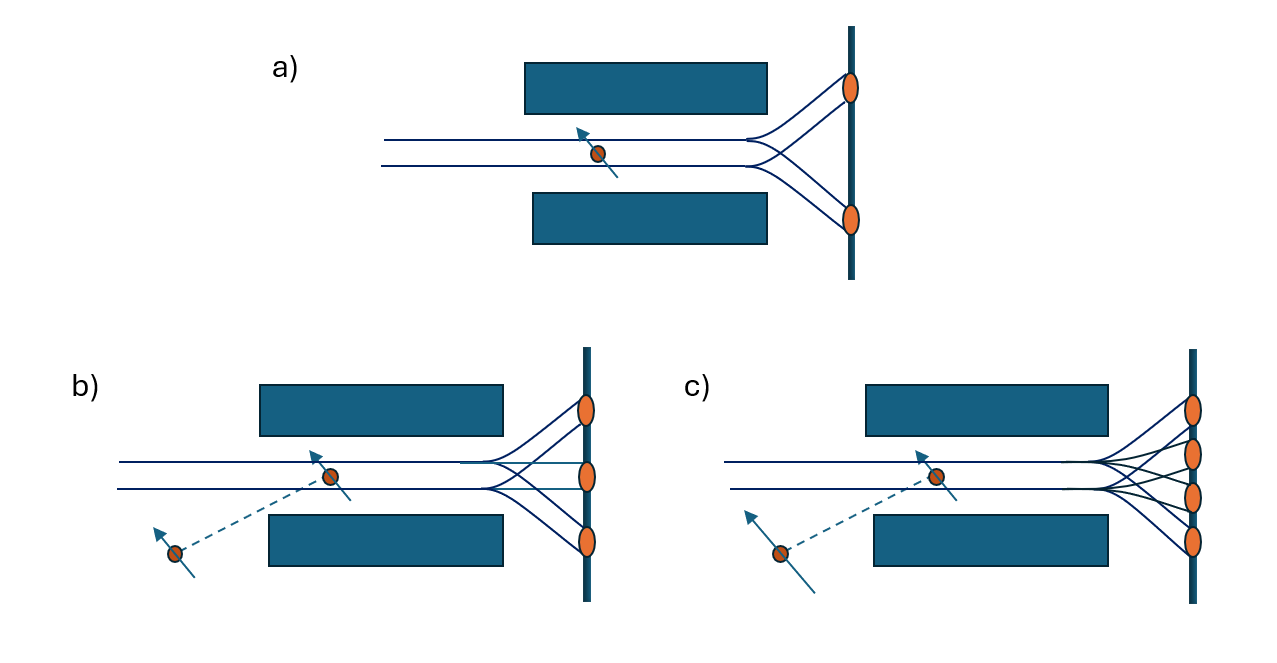}
    \caption{ a) A standard Stern-Gerlach experiment for a spin-half particle. b) The same Stern-Gerlach experiment, but the spin passing through the device interacts with an external spin-half particle which remains outside the device and constrains the spins to be parallel. Note that the outer spots in the second figure are in same positions as in the first figure. c) Similar to figure b), but the spin-half particle inside the device is interacting with a spin-one particle outside. }
    \label{fig:Fig1}
\end{figure}

To achieve this, we will imagine that the spin passing through the measuring device is coupled to another spin, which remains outside the device\footnote{ Alternatively, we could imagine that the second particle has no magnetic moment so does not interact with the device; and indeed therefore such a second particle could be part of a composite particle which passes through the Stern-Gerlach device (rather than remaining outside as depicted in Figure \ref{fig:Fig1}).}, in such a way that the spins are forced to remain parallel. Even though only one particle interacts with the measuring device, the measured particle will behave as if it has higher spin, as shown in Figure \ref{fig:Fig1}. 

In order to prove these results, we also provide some simple non-perturbative bounds (i.e. bounds which are strictly obeyed rather than approximations which hold for a particular order of perturbation theory)  for the evolution of a system subjected to a strong constraint.

\section{Setup}

Consider a spin-half particle ($s_1=\frac{1}{2}$),  coupled to a second particle which can have any spin $s_2$, via the Hamiltonian 
\be
H_0 = - \beta \vs_1 \cdot \vs_2  \label{spin-interaction} 
\ee
where $\vs_1$ and $\vs_2$ are spin operators for the respective particles, and $\beta$ is a  positive constant. In this case, the eigenstates of the Hamiltonian separate into two degenerate subspaces corresponding to the magnitude of their combined spin. We will assume that the initial state of the spins are in the lowest energy subspace with total spin $s=s_2 + \frac{1}{2}$, which intuitively corresponds to the two spins being parallel. 

We now imagine that the first spin alone passes through a Stern-Gerlach experiment, which measures the $z$-component of the first particle's spin. We can model this by incorporating the $z$-momentum of the first particle as an additional variable, which is acted on by the inhomogeneous magnetic field in the device. Although the details of an actual Stern-Gerlach experiment are somewhat more complicated\footnote{in particular, we omit here the effects of the average magnetic field, the $x$- and $y$- components of the field, and the free evolution of the particle during the measurement (apart from carrying it through the device), as these complicate the analysis without significant effect}, this interaction can be effectively modeled via the Hamiltonian 
\be 
\Hint = -\gamma_1 S_{1,z} Z \frac{d B_z}{d z},
\ee
where $\gamma_1$ is the gyromagnetic ratio of the first spin, $Z$ is the $z$-position operator for the particle, and $\frac{d B_z}{d z}$ is the magnetic field gradient in the device, which is assumed to be constant throughout the region through which the particle passes. We will assume that the particle interacts with the device for a time $T$, after which it travels to a screen where its position will be approximately proportional to the $z$-momentum of the particle. This models the Stern-Gerlach device as implementing a von Neumann measurement of $S_{1,z}$, with the $z$-momentum of the particle playing the role of the measurement pointer. Indeed, the interaction potential generates a force on the particle in the $z$-direction proportional to its $z$-spin. 

In the usual situation in which a spin-$\frac{1}{2}$ particle is not interacting with anything else, this would lead to two distinct shifts in the momentum of the particle yielding two spots on a screen behind the device. However, we will show below that due to the interaction with the spin outside the device, the spin-$\frac{1}{2}$ behaves effectively as a particle of higher spin $s=s_2 + \frac{1}{2}$, giving rise to more spots on the screen. In particular, we will show that when the coupling between the spins is sufficiently strong, such that they remain approximately parallel throughout the experiment, the effective coupling to the Stern-Gerlach device will be given by 
\be
V_{SG} \approx 
-\gamma^{\eff}_1 S_{1,z}^{\eff} Z  \frac{d B_z}{d z},
\ee
where $S_{1,z}^{\eff}$ is a spin operator for spin ($s_2 + \frac{1}{2}$), and $\gamma_1^{\eff}$ is a smaller gyromagnetic ratio that we calculate below. 

The full Hamiltonian for the system is given by $H_0 + V_{SG}$. We show in Appendix A that 
\be
H_0 = E_0 \Pi_s + c
\ee
where $\Pi_s$ is the projector onto the total spin $s=s_2 + \frac{1}{2}$ subspace, $E_0=-\beta \hbar^2(s_2 + \frac{1}{2})$, and $c=\frac{\beta}{2} \hbar^2 (s_2+1)$. Eliminating the  constant energy shift $c$, which has no physical effect, we will represent the dynamics of the system by the slightly simpler Hamiltonian 
\be
H_{1}^{SG}= E_0 \Pi_s + \Hint.
\ee

\section{Effective evolution}

When $|E_0|$ is sufficiently large relative to the interaction strength, we can prove two key results about the evolution generated by such a Hamiltonian. Firstly, that any state initially in the subspace spanned by $\Pi_s$ will remain contained within it with high probability for all time (Result 1.1 below). Secondly, that the time evolution for a substantial duration will be close to that generated by 
\be
H^{SG}_{2}= E_0 \Pi_s + \Pi_s \Hint \Pi_s,
\ee
in which the interaction is replaced by its projection into the subspace (Result 1.2 below). Result 1.1 is similar to the  general leakage bounds obtained in \cite{Szabo25}, using the Bloch \cite{Bloch58} and Schrieffer-Wolff \cite{Schrieffer66, Brayvi11} effective Hamiltonians. Result 1.2 is similar to the universal error bound obtained in \cite{gong20a, gong20b}. In both cases we offer alternative streamlined proofs for completeness in the Appendix, designed for our specific setup. For Result 1.1, this yields a tighter bound than the previous general results.  

In particular we prove the following:

\begin{result}
 Consider a Hamiltonian
\begin{equation}
    H_1=E_0\Pi +V,
\end{equation}
where $\Pi$ is an arbitrary projection operator, $V$ an arbitrary interaction Hamiltonian, and $E_0$ is a constant with dimensions of energy, which we will consider to be very large in magnitude (relative to the difference between the largest and smallest eigenvalue of $V$).  Consider a state $\ket {\Psi_1(0)}$ which is an eigenstate of $\Pi$ with eigenvalue 1. Then the following results hold
\begin{enumerate}
\item The state of the system has only a very small component outside the $+1$ eigenspace of $\Pi$  for all time. Namely the probability that the system is found outside the $+1$ eigenspace of $\Pi$ at time $t$ is \begin{equation}
 \mathrm{Prob}({\rm outside\ subspace},t) = \bra{\Psi_1(t)} I-\Pi \ket{\Psi_1(t)} \leq \left( \frac{\DeltaV}{E_0}\right)^2 \label{stay-in-subspace}
  \end{equation}
    where \[\ket {\Psi_1(t)} = U_1(t)\ket {\Psi_1(0)}=e^{-iH_1t/\hbar}\ket {\Psi_1(0)},\]
and $\DeltaV$ is the difference between the maximum and minimum eigenvalue\footnote{For a finite dimensional space in which $V$ has eigenvalues $\lambda_1 \leq \lambda_2 \leq \lambda_d$, then $\DeltaV = \lambda_d - \lambda_1$. More generally, for an infinite dimensional Hermitian operator $V$ with spectrum $\sigma(V)$, we can  define $\DeltaV = \sup(\sigma(V)) - \inf(\sigma(V))$.} of $V$, and $I$ is the identity operator.

    \item For any given time $T$, for sufficiently large $|E_0|$ the dynamics generated by $H_1$ during the interval $0 \leq t \leq T$ is close to that generated by 
    \begin{equation}
    H_2=E_0\Pi +\Pi V \Pi.
    \end{equation}
    In particular let  $\ket {\Psi_2(t)} $ be the state that evolves under $H_2$, starting with the same initial state, $\ket {\Psi_2(0)} =\ket {\Psi_1(0)} $, then the fidelity between the two states is given by 
    \be
    \left|\braket{\Psi_2(t)}{\Psi_1(t)}\right|^2 = 1 - \mu(t) 
    \ee
    where 
    \be
     \mu(t) \leq \frac{ t}{\hbar}\frac{(\DeltaV)^2}{|E_0|}. \label{eq:D_bound}
    \ee
Note that $\mu(t)$ can be made as small as desired over the entire time interval $0 \leq t \leq T$ by taking $|E_0|$ sufficiently large. 

\end{enumerate}

\end{result}

\begin{proof}
See Appendix B
\end{proof}

Applying the above result to the Stern-Gerlach experiment, we find that by taking $|E_0|$ sufficiently large the evolution of the particle will be well described by the Hamiltonian 
\be \label{eq:H2}
H_{2}^{SG} = E_0 \Pi_s -\gamma_1 (\Pi_s S_{1,z} \Pi_s) Z  \frac{d B_z}{d z}.
\ee
A slight complication arises in applying the result, because $Z$ is an unbounded operator and hence $\mathcal{D}(V_{SG})$ is infinite. However, we will consider an initial state in which the particle is be contained within a finite region near the centre of the Stern-Gerlach device (with $-z_0 \leq z \leq z_0$). As the evolution generated by $H^{SG}_1$ or $H^{SG}_2$ commutes with $Z$, it will remain in this region throughout its evolution through the Stern-Gerlach device. We can therefore replace the operator $Z$ by a bounded operator which is zero outside this region without affecting its evolution. This gives $\mathcal{D}(V_{SG})= \gamma_1 \hbar z_0 \frac{dB_z}{dz}$. It then follows from \eqref{eq:D_bound} that $H^{SG}_2$ will closely approximate the true evolution when 
\begin{equation}
\frac{ T}{\hbar}  \frac{(  \gamma_1 \hbar z_0 \frac{dB_z}{dz})^{2}}{|E_0|}  \ll 1, \label{eq:approximation}
\end{equation}
which is achievable for any values of the other parameters by taking $|E_0|$ sufficiently large.

Considering \eqref{eq:H2}, we note that $H_2^{SG}$ commutes with $\Pi_s$, hence the system will remain within the subspace spanned by $\Pi_s$ throughout the evolution generated by $H_2^{SG}$, while the Stern-Gerlach experiment essentially performs a measurement of $\Pi_s S_{1,z} \Pi_s$. 

\section{Effective spin}

We now come to the key result of our paper, which is that $\Pi_s S_{1,z} \Pi_s$ acts like a spin operator for a particle of higher spin (in particular of the total spin $\jmax$ rather than $s_1$). This leads to ($2\jmax+1$) spots in the Stern-Gerlach experiment, even though the device is only interacting with the spin-half particle (i.e. measuring the spin of the spin-half particle in the $z$-direction). However the effective gyromagnetic ratio of the particle is reduced. 

More precisely, we find 

\begin{result} The operators
\be \label{eq:Seff}
S_{1,k}^{\eff} = (2 \jmax) \Pi_s S_{1,k} \Pi_s 
\ee
act on states in the subspace spanned by $\Pi_s$ (i.e. those states with combined spin $\jmax$) exactly as spin $\jmax$ operators, obeying all of the required commutator and eigenvalue equations. 
\end{result}
\begin{proof}
   We give a simple proof\footnote{An alternative approach to proving this would be to use the Projection Theorem \cite{sakurai2014} obtained from the Wigner-Eckart Theorem.} of this in Appendix \ref{app:proof2}. 
\end{proof}

Defining an effective gyromagnetic ratio by 
\be
\gamma_1^{\eff} = \frac{\gamma_1}{2 \jmax}
\ee
we can use \eqref{eq:Seff} to rewrite $H_2$ as 
\be
H_2 = E_0 \Pi_s -\gamma^{\eff}_1 S_{1,z}^{\eff} Z  \frac{d B_z}{d z}.
\ee

Intuitively, the interaction term will induce a force on the particle proportional to $S_{1,z}^{\eff}$, which will cause a corresponding shift in the $z$-momentum of the particle. In particular, each eigenstate $\ket{\jmax, \mj}$ of $S_{1,z}^{\eff}$ will acquire a momentum shift proportional to its eigenvalue $\mj \hbar$. If these shifts separate the momenta for different eigenstates by much more than the spread in momentum in the initial state then the measurement will lead to $(2 \jmax +1)$ distinct spots on the screen after the Stern-Gerlach device.

We show this in detail for a simple model in which the initial wavefunction is a `top-hat' function in $z$ of width $2z_0$ in Appendix D. For this model the requirement that the spots can be easily distinguished is given by 
\be 
\gamma_1^{\eff} T \frac{dB_z}{dz} \gg \frac{\pi}{z_0}.
\ee

An interesting observation is that the total range of momentum shifts generated by the Stern-Gerlach device via $H_2$ lies in the same range (between $-\gamma_1 \frac{\hbar}{2}\frac{dB_z}{dz} $ and $+\gamma_1 \frac{\hbar}{2}\frac{dB_z}{dz} $) as the momentum shifts which would be expected for a standard spin-half particle interacting with the device. This is because the increased range of spin eigenvalues is compensated for by the reduced effective gyromagnetic ratio. Hence the upper and lower spot will always align with those we would expect for a single spin-half particle. 

Note that it is possible to prepare an initial state such that the spin-half particle in the Stern-Gerlach device deterministically reaches any particular spot on the screen\footnote{This is different from a generalised measurement described by a POVM (positive operator valued measure) on the spin-half particle, which could produce more than two outcomes, but with no way to achieve each outcome deterministically.} (by preparing the joint state of the two spins as $\ket{\jmax, \mj}$).

\section{Discussion}
We have shown that a spin $\frac{1}{2}$ particle, which is strongly coupled to another spin, can behave under interactions that only affect itself as if it were larger spin. The Stern-Gerlach experiment is a striking example of this, where the spin-half particle yields more than two distinct spots when its spin is measured.  Crucially we note that for any chosen spot on the screen, an individual spin can be prepared in a state, so that with probability one it goes to that spot. It is interesting to speculate about the birth of quantum theory at the start of the 20th century, had such spin couplings been generic in nature.  The original Stern-Gerlach experiment would have yielded very different results, which may have significantly hindered our understanding of quantum spin. 

More broadly, given a sufficiently strong coupling to another spin,  a spin $\frac{1}{2}$ particle will behave as a larger spin in all scenarios, albeit with a reduced gyromagnetic ratio. For example, if the second spin were also spin half, $s_2=\frac{1}{2}$ (and hence the spin-$\frac{1}{2}$ particle behaves as if it had spin 1), and we interacted the spin-$\frac{1}{2}$ particle with a different spin 1 particle, we would expect it to be able to flip the other spin while maintaining total angular momentum. 

Furthermore, similar phenomena may play a role in other scenarios. For example, we could consider a hypothetical model of a solid state system as illustrated in Figure \ref{solid state fig}.
\begin{figure}
    \centering \label{solid state fig} 
    \includegraphics[width=10cm]{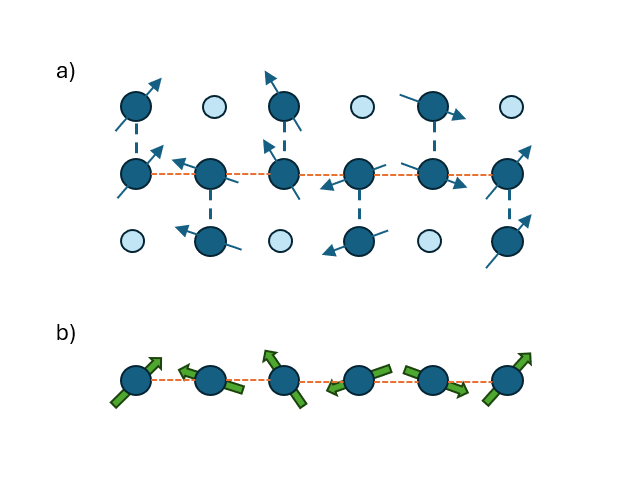}
    \caption{ a) shows a hypothetical three-layer material composed of spin $\frac{1}{2}$ particles (in dark blue with arrows) and ``inert'' particles (in light blue). There are strong inter-layer interactions which constrain the spins to remain parallel, and a weaker arbitrary interaction between spins in the middle layer. b) shows an effective model of the middle layer as coupled spin 1 particles.  }
    \label{solid state fig}
    \label{fig:placeholder}
\end{figure}
The material is composed of three layers; the middle layer contains spin $\frac{1}{2}$ particles at each lattice site; whilst the upper and lower layers contain spin $\frac{1}{2}$ particles at every second lattice site and  ``inert" particles in between which do not interact with the spins. The upper and lower levels are shifted by one lattice site relative to each other, so that each of the middle layer spins is in the vicinity of a single spin from the outer layers.  All of the interactions are nearest-neighbour. Let the spins in the middle layer interact with the outer layer spins via a strong spin-spin  interaction similar to $H_0$ in \eqref{spin-interaction}. The interaction between the spins of the middle layer is much weaker than that in between the spins of the middle layer and the outer spins. In this situation we may imagine a phase in which the middle layer spins are parallel with the outer spins. They then behave as spin-1 particles as far as the mid-layer interactions are concerned. It is known that spin-$1$ chains have markedly different properties than spin-$\frac{1}{2}$ chains \cite{haldane83a, Haldane83b,Affleck89}. This idea could also be extended by considering different arrangements of spins above and below the central line, to construct more exotic spin chains with higher effective spin.

A related example can be envisaged in molecular systems. For example a linear chain molecule ABBA where the A-B interaction is much stronger than the B-B interaction. The B atoms will behave as if they have higher spin or angular momentum than they naturally have.

Finally, in the present paper we have discussed the particular case of spins, but we would expect analogous mechanisms to apply to any scenario in which we are interacting with only one part of a strongly coupled system. The part we are interacting with may behave as if it has very different properties from the original system, including the  dimension of its Hilbert space.

\bibliography{mybib}{}
\bibliographystyle{ieeetr}

\section*{Appendices}
\appendix

\section{Coupled spins}
Note that we can rewrite the initial Hamiltonian as 
\begin{align}
H_0 &= - \beta \vs_1 \cdot \vs_2 \nonumber \\
& = - \frac{\beta}{2} \left( J^2 - S_1^2 - S_2^2 \right),
\end{align}
where we have denoted the total spin by $\underline{J}= \vs_1 + \vs_2$. Following standard procedures for combining angular momentum, a complete set of eigenstates for this Hamiltonian are the combined spin states $\ket{j, \mj}$ satisfying 
\begin{align} 
J_z \ket{j, \mj} &= \mj \hbar \ket{j, \mj} \\
J^2\ket{j, \mj} &= j(j+1) \hbar^2\ket{j, \mj}\\
S_1^2\ket{j, \mj} &=  \frac{3}{4}\hbar^2\ket{j, \mj}\\
S_2^2\ket{j, \mj} &=  s_2(s_2+1) \hbar^2\ket{j, \mj}
\end{align} 
where the quantum numbers satisfy $j\in \{\jmin, \jmax \}$, where $s=s_2 + \frac{1}{2}$, and $\mj \in \{ -j, -j+1, \ldots, j\}$. It follows that 
\begin{align} 
H_0\ket{\jmax, \mj} &= -\frac{\beta \hbar^2 s_2}{2}\ket{\jmax, \mj} \\
H_0\ket{\jmin, \mj} &=  \frac{\beta \hbar^2}{2} \left( s_2 +1  \right) \ket{\jmin, \mj}.
\end{align} 
Hence  
\be
H_0 = - \beta \hbar^2 \left(s_2 + \frac{1}{2}\right) \Pi_s + \frac{\beta\hbar^2}{2}  (s_2 + 1) 
\ee
 where 
\be
\Pi_s = \sum_{\mj = -\jmax}^{\jmax} \proj{\jmax,\mj }.
\ee
Incorporating the interaction with the Stern-Gerlach device, and eliminating the constant energy shift, we obtain the Hamiltonian 
\be
H_1 = E_0 \Pi_s + V_{\mathrm{SG}}
\ee
as required, where $E_0=- \beta \hbar^2 \left(s_2 + \frac{1}{2}\right)$. 

\section{Proof of Result 1}
Consider a Hamiltonian
\begin{equation}
    H_1=E_0\Pi +V,
\end{equation}
where $\Pi$ is an arbitrary projection operator and $V$ an arbitrary interaction Hamiltonian; $E_0$ is a constant with dimensions of energy, which we will consider to be very large in magnitude (relative to the difference between the maximum and minimum eigenvalue of $V$).  Consider a state $\ket {\Psi_1(0)}$ which is an eigenstate of $\Pi$ with eigenvalue 1.

Recall that $\DeltaV$ denotes the difference between the maximum and minimum eigenvalues of $V$. Below we consider the case in which these eigenvalues are arranged symmetrically about the origin, for which $\DeltaV = 2 \| V\|$, where $\|V\|=\sup \{ \| V \ket{\psi} \| : \|\ket{\psi}\| = 1, \ket{\psi} \in \mathcal{H}\}$ denotes the operator norm of $V$. If this is not the case, we can always shift $V$ by a constant without affecting the dynamics, to obtain the same bound expressed in terms of $\DeltaV$.

\begin{enumerate}
    \item The state of the system has only a very small component outside the $+1$ eigenspace of $\Pi$  for all time.
\begin{equation}
 \mathrm{Prob}({\rm outside\ subspace},t) = \bra{\Psi_1(t)} I-\Pi \ket{\Psi_1(t)} \leq \left(\frac{\DeltaV}{E_0}\right)^2 \label{stay-in-subspace}
  \end{equation}
    where \[\ket {\Psi_1(t)} = U_1(t)\ket {\Psi_1(0)}=e^{-iH_1t/\hbar}\ket {\Psi_1(0)}.\]
\end{enumerate}
{\em Proof}
\begin{eqnarray}\bra{\Psi_1(t)}I-\Pi\ket {\Psi_1(t)} &=&   \bra{\Psi_1(t)}(I-\Pi)^2 \ket {\Psi_1(t)} \nonumber\\
&=& \bra{\Psi_1(t)}\left( I-\frac{H_1}{E_0} +\frac{V}{E_0} \right)^2\ket {\Psi_1(t)} \nonumber\\
&=& \bra{\Psi_1(t)}\left(\left(I-\frac{H_1}{E_0}\right)^2 +\left(I-\frac{H_1}{E_0}\right) \frac{V}{E_0} + \frac{V}{E_0}\left(I-\frac{H_1}{E_0}\right) + \frac{V^2}{E_0^2} \right)\ket {\Psi_1(t)} \nonumber\\
&=& \bra{\Psi_1(0)}\left(\left(I-\frac{H_1}{E_0}\right)^2 +\left(I-\frac{H_1}{E_0}\right) \frac{U_1(t)^\dagger V U_1(t)}{E_0}\right.\nonumber \\ & & \qquad\qquad  \left. + \frac{U_1(t)^\dagger V U_1(t)}{E_0}\left(I-\frac{H_1}{E_0}\right)  
 + \frac{\left(U_1(t)^\dagger V U_1(t)\right)^2}{E_0^2} \right)\ket {\Psi_1(0)} \nonumber\\
&=&  \bra{\Psi_1(0)}\left(\frac{V^2}{E_0^2} -\frac{V  \left(U_1(t)^\dagger V U_1(t)\right)}{E_0^2}\right. \nonumber \\
& & \qquad \qquad  \left. - \frac{\left(U_1(t)^\dagger V U_1(t)\right) V}{E_0^2}+ \frac{\left(U_1(t)^\dagger V U_1(t)\right)^2}{E_0^2}  \right)\ket {\Psi_1(0)} \nonumber\\
&=& \bra{\Psi_1(0)}\left(\frac{V}{E_0} - \frac{ U_1(t)^\dagger V U_1(t)}{E_0}\right)^2\ket {\Psi_1(0)} \nonumber\\
&=& \left\|\left(\frac{V}{E_0} - \frac{ U_1(t)^\dagger V U_1(t)}{E_0}\right)\ket {\Psi_1(0)} \right\|^2 \nonumber \\
&\leq & \left\|\left(\frac{V}{E_0} - \frac{ U_1(t)^\dagger V U_1(t)}{E_0}\right) \right\|^2 \nonumber \\
&\leq & \left( \left\|\frac{V}{E_0} \right\| + \left\| \frac{U_1(t)^\dagger V U_1(t)}{E_0} \right\| \right)^2  \nonumber \\
&= & \left( \frac{2||V||}{|E_0|}\right)^2 \nonumber \\
&=& \left(\frac{\DeltaV}{E_0}\right)^2, \qquad \qquad\qquad \qquad\square
\end{eqnarray}
where in the fourth line we have used the fact that $\left(I-\frac{H_1}{E_0}\right)$ commutes with $ U_1(t)$,  in the fifth line we have used the fact that $\left(I-\frac{H_1}{E_0}\right)\ket{\Psi_1(0)} = -\frac{V}{E_0}\ket{\Psi_1(0)}$, and in the eighth line we have moved from the vector norm to the operator norm.

\begin{enumerate}\setcounter{enumi}{1} 

\item Consider now a second Hamiltonian
\begin{equation}
    H_2=E_0\Pi +\Pi V\Pi,
\end{equation}
and let $\ket {\Psi_2(t)} $ be the state that evolves under $H_2$, starting with the same initial state, $\ket {\Psi_2(0)} =\ket {\Psi_1(0)} $.  And let \footnote{Note that the trace distance between two pure states is given by $d(\ket{\Psi_1}, \ket{\Psi_2}) = \sqrt{1-|\braket{\Psi_2}{\Psi_1}|^2}$. Hence, we can also use this result to bound the trace distance via  $d(\ket{\Psi_1(t)}, \ket{\Psi_2(t)})= \sqrt{\mu(t)}$.}

\be \mu(t):=1-\left| \langle {\Psi_2(t)} \ket {\Psi_1(t)}\right|^2 .  \ee

    Then
   \begin{equation}
     \mu(t) \leq \frac{ t}{\hbar}\frac{(\DeltaV)^2}{|E_0|}. \end{equation} 
     Note that this implies that for any fixed time interval $0 \leq t \leq T$, for sufficiently large $|E_0|$, the dynamics under $H_1$ and $H_2$ are close:
\end{enumerate}
{\em Proof}
\begin{eqnarray} \mu(t)&=&1- \langle {\Psi_2(t)} \ket {\Psi_1(t)} \braket{\Psi_1(t)}{\Psi_2(t)}  \nonumber\\
\end{eqnarray}
So
\begin{eqnarray} \frac {d\mu(t)}{dt}
&=&-\bra {\Psi_2(t)} \left(\frac{iH_2}{\hbar}-\frac{iH_1}{\hbar}\right)\ket {\Psi_1(t)}\braket{\Psi_1(t)}{\Psi_2(t)}  -  \braket{\Psi_2(t)}{\Psi_1(t)}\bra {\Psi_1(t)} \left(\frac{iH_1}{\hbar}-\frac{iH_2}{\hbar}\right)\ket {\Psi_2(t)} \nonumber\\
&=&\frac{i}{\hbar}\Big(\bra{\Psi_2(t)} \left(V- \Pi V \Pi\right)\ket {\Psi_1(t)}\braket{\Psi_1(t)}{\Psi_2(t)}  - \braket{\Psi_2(t)}{\Psi_1(t)}\bra {\Psi_1(t)} \left(V- \Pi V \Pi\right)\ket {\Psi_2(t)}\Big) \nonumber\\
&=&\frac{i}{\hbar}\Big(\bra{\Psi_2(t)} V \left(I- \Pi \right)\ket {\Psi_1(t)}\braket{\Psi_1(t)}{\Psi_2(t)}  - \braket{\Psi_2(t)}{\Psi_1(t)}\bra {\Psi_1(t)} \left(I- \Pi\right) V \ket {\Psi_2(t)}\Big), 
\nonumber\\
\end{eqnarray}
where in the first line we have used the Schrodinger equations 
$i \hbar \frac{d}{dt} \ket{\Psi_k(t)} = H_k \ket{\Psi_k(t)}$ for $k \in \{1,2\}$ and their Hermitian conjugates, and in the
third line we have used the fact that $\Pi \ket{\Psi_2(t)} = \ket{\Psi_2(t)}$, because $[\Pi, H_2]=0$, and $\Pi \ket{\Psi_2(0)} =  \ket{\Psi_2(0)}$.

Thus using Cauchy-Schwarz and triangle inequalities,
\begin{eqnarray} \left|\frac {d\mu(t)}{dt} \right|
&\leq&
\frac{2}{\hbar}\, \|V \ket {\Psi_2(t)}\|\, \| (I-\Pi) \ket {\Psi_1(t)}\| \left|\braket{\Psi_1(t)}{\Psi_2(t)} \right| \nonumber\\
&\leq&\frac{2}{\hbar}\, \|V \|\, \sqrt{\bra{\Psi_1(t)} (I-\Pi)  \ket {\Psi_1(t)}} \nonumber\\
&\leq&
\frac{2}{\hbar}\, \|V \|\,\frac{2 \| V\|}{|E_0|}\nonumber\\
&=& \frac{1}{\hbar} \frac{(\DeltaV)^2}{|E_0|},
\end{eqnarray}
where in the third line we have used \eqref{stay-in-subspace}, and in the fourth line we have used $\DeltaV=2 \| V\|$.

Given that $\mu(0)=0$, and we have bounded the derivative $\frac {d\mu(t)}{dt}$, we therefore obtain 
   \begin{equation}
       \mu(t) \leq \frac{t}{\hbar}\frac{\left(\DeltaV\right)^2}{|E_0|}.\qquad\qquad\qquad\qquad\square
   \end{equation}

Note that unlike the probability to be found outside the subspace spanned by $\Pi$, which remains small for all times, $\mu(t)$ may become large over time. This can be seen by considering  a three dimensional quantum system with orthonormal basis $\ket{0}, \ket{1}, \ket{2}$, with $\Pi = \proj{0} + \proj{1}$ and $V= \epsilon (\ketbra{1}{2} + \ketbra{2}{1})$ where $0<\epsilon \ll E_0$, and the initial state $\ket{\Psi_1(0)}= \frac{1}{\sqrt{2}} \left( \ket{0} + \ket{1} \right)$. At time $t= \frac{\pi \hbar E_0}{\epsilon^2}$ the states $\ket{\Psi_1(t)}$ and $\ket{\Psi_2(t)}$ become approximately orthogonal (and thus $\mu(t) \approx 1$).

\section{Proof of Result 2} \label{app:proof2}
In this Appendix, we show that the operators given by 
\be
S_{1,k}^{\eff} := (2 \jmax) \Pi_s S_{1,k} \Pi_s
\ee
for $k \in \{x,y,z\}$ act on the $(2 \jmax +1)$-dimensional subspace of states spanned by $\Pi_s$ as if they were spin operators with spin $\jmax$. We will show this by proving that $S_{1,z}^{\eff}$ and the raising and lowering operators $S_{1,+}^{\eff}$ and $S_{1,-}^{\eff}$ act appropriately for spin $\jmax$.

We first consider the action of the operator $S_{1,z}^{\eff}$
on the states $\ket{\jmax,\mj}$. In order to do this it will be helpful to decompose the combined spin states in the individual spin basis given by $\ket{m_1, m_2}$, corresponding to the $z$-components of spin for the two particles. Using standard Clebsch-Gordan coefficients \cite{CondonShortley1935}, we find 
\be 
\ket{\jmax,\mj}  = \sqrt{\frac{1}{2} \left( 1 + \frac{\mj}{\jmax} \right) }\ket{\frac{1}{2}, \mj-\frac{1}{2}} + \sqrt{\frac{1}{2} \left( 1 - \frac{\mj}{\jmax} \right) }\ket{-\frac{1}{2}, \mj+\frac{1}{2}}.
\ee
Now
\begin{align}
S_{1,z}^{\eff} \ket{\jmax,\mj} & = (2 \jmax)\Pi_s S_{z,1} \Pi_s \ket{\jmax,\mj} \nonumber \\
&= (2 \jmax)\Pi_s S_{z,1} \ket{\jmax,\mj} \nonumber \\
&= (\jmax \hbar) \, \Pi_s\,  \left( \sqrt{\frac{1}{2} \left( 1 + \frac{\mj}{\jmax} \right) }\ket{\frac{1}{2}, \mj-\frac{1}{2}} - \sqrt{\frac{1}{2} \left( 1 - \frac{\mj}{\jmax} \right) }\ket{-\frac{1}{2}, \mj+\frac{1}{2}} \right) \nonumber\\
& = (\jmax \hbar) \, \proj{\jmax, \mj} \left( \sqrt{\frac{1}{2} \left( 1 + \frac{\mj}{\jmax} \right) }\ket{\frac{1}{2}, \mj-\frac{1}{2}} - \sqrt{\frac{1}{2} \left( 1 - \frac{\mj}{\jmax} \right) }\ket{-\frac{1}{2}, \mj+\frac{1}{2}} \right) \nonumber\\
& =  (\jmax \hbar) \left[ \frac{1}{2} \left( 1 + \frac{\mj}{\jmax}\right)- \frac{1}{2} \left( 1 - \frac{\mj}{\jmax} \right) \right]\ket{\jmax,\mj} \nonumber \\
& = \mj \hbar \ket{\jmax,\mj},
\end{align}
where in the second line we have used the fact that $\Pi_s\ket{\jmax,\mj} = \ket{\jmax,\mj}$, and in the fourth line we have used the fact that in the projector 
\be
\Pi_s = \sum_{\mj'=-\jmax}^{\jmax} \proj{\jmax, \mj'}
\ee
 the only term with non-zero contribution is the one in which $\mj'=\mj$. Hence the operator $S_{1,z}^{\eff}$ acts on the states $\ket{\jmax,\mj}$ exactly as if it were the operator for the $z$-component of spin of a standard spin $\jmax$ particle. 

 We also want to show that the operators $S_{1,x}^{\eff}$ and $S_{1,y}^{\eff}$ act on the states $\ket{\jmax,\mj}$ in the same way as standard spin operators in the $x$- and $y$-direction. It is equivalent and  more convenient to show that the raising and lowering operators 
\begin{align}
S_{1,+}^{\eff} &=S_{1,x}^{\eff} + i S_{1,y}^{\eff} \\
S_{1,-}^{\eff} &=S_{1,x}^{\eff} - i S_{1,y}^{\eff}
\end{align}
act appropriately. In particular 
\begin{align} 
S_{1,+}^{\eff} \ket{\jmax,\mj} &=  (2 \jmax)\Pi_s S_{1,+} \Pi_s \ket{\jmax,\mj} \nonumber \\
&= (2 \jmax)\Pi_s S_{1,+} \ket{\jmax,\mj} \nonumber \\
&= (2 \jmax \hbar)\Pi_s  \sqrt{\frac{1}{2}\left(1-\frac{\mj}{\jmax}\right)} \ket{\frac{1}{2}, \mj +\frac{1}{2} } \nonumber \\
&= (2 \jmax \hbar ) \proj{\jmax, \mj + 1} \left(\sqrt{\frac{1}{2}\left(1-\frac{\mj}{\jmax}\right)}  \ket{\frac{1}{2}, \mj +\frac{1}{2} } \right)\nonumber \\
&= (2 \jmax \hbar )\sqrt{\frac{1}{2}\left(1+\frac{\mj+1}{\jmax}\right)}\sqrt{\frac{1}{2} \left(1-\frac{\mj}{\jmax}\right)} \ket{\jmax, \mj+1} \nonumber \\
&= \hbar \sqrt{(\jmax + \mj + 1)(\jmax - \mj)}\ket{\jmax, \mj+1}, \label{raising}
\end{align}
where in the third line we have used the spin-half relations $S_+\ket{-\frac{1}{2}}= \hbar \ket{\frac{1}{2}}, S_+\ket{\frac{1}{2}}=0.$ Note that \eqref{raising} is the standard relation for a spin $\jmax$ raising operator. Similarly 
\begin{align} 
S_{1,-}^{\eff} \ket{\jmax,\mj} &=  (2 \jmax)\Pi_s S_{1,-} \Pi_s \ket{\jmax,\mj} \nonumber \\
&= (2 \jmax)\Pi_s S_{1,-} \ket{\jmax,\mj} \nonumber \\
&= (2 \jmax \hbar)\Pi_s \sqrt{\frac{1}{2} \left( 1 + \frac{\mj}{\jmax} \right) }\ket{-\frac{1}{2}, \mj-\frac{1}{2}} \nonumber \\
&= (2 \jmax \hbar ) \proj{\jmax, \mj - 1} \left(\sqrt{\frac{1}{2}\left(1+\frac{\mj}{\jmax}\right)}  \ket{-\frac{1}{2}, \mj -\frac{1}{2} } \right)\nonumber \\
&= (2 \jmax \hbar )\sqrt{\frac{1}{2} \left( 1 - \frac{\mj-1}{\jmax} \right) }\sqrt{\frac{1}{2} \left(1+\frac{\mj}{\jmax}\right)} \ket{\jmax, \mj-1} \nonumber \\
&= \hbar \sqrt{(\jmax - \mj + 1)(\jmax + \mj)}\ket{\jmax, \mj-1},
\end{align}
which is the standard relation for a lowering operator for a spin $\jmax$ particle.

It follows that the effective spin operators $S_{1,k}^{\eff}$ obey all of the usual angular momentum relations when acting on the states $\ket{\jmax, \mj}$. In particular 
\begin{align} 
[S_{1,i}^{\eff}, S_{1,j}^{\eff} ] &= i \hbar \epsilon_{ijk} S_{1,k}^{\eff}, \\
S_{1,z}^{\eff} \ket{\jmax, \mj} &= \mj \hbar\ket{\jmax, \mj}, \\
\left((S_{1,x}^{\eff})^2 + (S_{1,y}^{\eff})^2 + (S_{1,z}^{\eff})^2 \right)\ket{\jmax, \mj} &= \jmax (\jmax+1) \hbar^2 \ket{\jmax, \mj},
\end{align}
where $\epsilon_{ijk}$ is the Levi-Civita symbol. In the first equation we have used the fact that all of the effective spin operators act like the 0 operator outside of  the spin $\jmax$ subspace so the commutation relations are satisfied for all states, and not just those within the subspace.

\section{Measurement details}
   
We now consider how the previous results apply to the measurement process.
Following the von Neumann approach to measurement, we consider a system coupled to a pointer (represented by a particle on a line, the momentum $P$ of which will indicate the measurement result).

We evolve the system and measuring device together for a time $T$ under the Hamiltonian
\begin{equation}
H_3=E_0 \tilde{\Pi} \otimes I - R\otimes Z\end{equation}
where $R$ is a bounded operator on the system, which is to be measured,  $\tilde{\Pi} $ is a projection operator on the system, and  $Z$ is the position operator of the pointer in the $z$-direction. Note that here we have included explicit tensor products for clarity. 

In order to use our previous results which required bounded operators, we consider an initial pointer state  $\ket{\psi(0)}$ which  has support only within a finite region $-z_0\leq z\leq z_0$. As  $Z$ commutes with $H_3$, the pointer will remain within this region throughout the evolution, allowing us to replace the position operator $Z$ by the bounded operator
\begin{equation}
    \hatZ = \int_{-z_0}^{z_0} dz\ z \ket z\bra z
\end{equation} 
without affecting the dynamics. For any pointer state of this type, the evolution under $H_3$ is exactly the same as that under 
\begin{equation}
H_4=E_0 \tilde{\Pi} \otimes I - R\otimes \hatZ\end{equation}

Taking $\Pi = \tilde{\Pi} \otimes I$ and $V = -R\otimes \hatZ$, we note that $\DeltaV=2z_0\| R\|  $. Our previous results imply that for sufficiently large $E_0$, satisfying 
\begin{equation}
\frac{ T}{\hbar} \frac{(2z_0\| R\|)^2 }{|E_0|} \ll 1, \label{eq:approximation}
\end{equation}
the  evolution under the Hamiltonian $H_4$ for time $T$ is approximately equivalent to that under 
\begin{equation}
H_5=E_0 \tilde{\Pi} \otimes I - \tilde{\Pi} R \tilde{\Pi} \otimes \hatZ,
\end{equation}

Now let us consider a specific choice of pointer state at $t=0$: it is a top hat function in position space:
\begin{equation}
\ket{\psi(0)} =\int_{-z_0}^{z_0}  dz  \frac{1}{\sqrt{2z_0}} \ket {z} = \int_{-\infty}^{\infty}  dp\, \sqrt{\frac{z_0}{\pi \hbar}} 
\,\frac{\sin(p z_0/\hbar)}{pz_0/\hbar}\ket{p}, 
\label{pointer-state}\end{equation}
which corresponds to a sinc function when represented in terms of momentum states $\ket{p}$,
\begin{equation}
\ket{p}  = \frac{1}{\sqrt{2\pi\hbar}}\int_{-\infty}^{\infty}  dz\, e^{ipz/\hbar} 
\,\ket{z}. 
\end{equation}

We will also consider the case in which the system is in an eigenstate of $\tilde{\Pi} R \tilde{\Pi}$ with eigenvalue $\lambda$, and of $\tilde{\Pi}$ with eigenvalue 1. Then the final state of the pointer when evolved via $H_5$ will be shifted in momentum\footnote{The combined evolution of the  system and pointer is given by $\ket{\Psi(T)} = e^{-\frac{i}{\hbar}H_5 T} \ket{\lambda} \ket{\psi(0)} = e^{-\frac{i}{\hbar} E_0 T}\ket{\lambda} \ket{\psi(T)}$.} to give
\begin{equation}
    \ket{\psi(T)}=\int_{-z_0}^{z_0}  dz  \frac{1}{\sqrt{2z_0}} e^{\frac{i}{\hbar} \lambda z T} \ket {z} =\int_{-\infty}^{\infty}  dp\,\sqrt{\frac{z_0}{\pi \hbar}}\, \frac{\sin((p - \lambda T) z_0/\hbar)}{(p - \lambda  T) z_0/\hbar} \ket{p}.
\end{equation}

Consider now that the minimum spacing between the eigenvalues of $\tilde{\Pi} R \tilde{\Pi}$ is $\delta_\lambda$. What we require is that their associated momentum-space wave-functions should be sufficiently distinguishable. The width of the momentum-space wave-functions are  $\pi\hbar/z_0$ (by this we mean the distance from the maximum of the function to the first zero).  So as long as we take the distance between the peaks to be much larger than this then the different eigenstates will be distinguishable with high probability.
i.e, if we choose 
\begin{equation}
\delta_\lambda T \gg \frac{\pi\hbar}{z_0} \label{eq:measurment_result}
\end{equation}

In particular, if the eigenvalues are equally spaced, and we take
$\delta_\lambda T = \frac{k\pi\hbar}{z_0}$, for a chosen constant $k$,
then we can associate measurement results of $p$ with eigenvalues $\lambda$ by dividing the momentum space into equal `bins' of width $\frac{k \pi \hbar}{z_0}$ centred on $\lambda T$. 
The probability of correctly identifying an eigenvalue if the evolution was exactly given by $H_5$ is 
\begin{equation}
    \int_{-\frac{k\pi\hbar}{2z_0}}^{+\frac{k\pi\hbar}{2z_0}} dp\left(\frac{z_0}{\pi\hbar}\right) \left(\frac{\sin(pz_0/\hbar)}{pz_0/\hbar}\right)^2 = \frac{1}{\pi}\int_{-k\pi/2}^{+k\pi/2}dq \left(\frac{\sin q}{q}\right)^2,
\end{equation}
(note this independent of any of the physical parameters of the problem). For sufficiently large $k$ we can make this probability as close to 1 as desired. For example for $k=10$, this integral is approximately 0.98.

In summary, our measuring device will provide an accurate measurement of $\tilde{\Pi} R \tilde{\Pi}$ as long as equations \eqref{eq:approximation} and \eqref{eq:measurment_result} are satisfied, which can always be achieved by choosing appropriate parameters for the measuring device and then taking $|E_0|$ sufficiently large.

\end{document}